\documentclass[12pt]{emulateapj}
\usepackage{natbib}
\usepackage{rotating}
\usepackage{mathptmx}
\newcommand{\etal}{et~al.}

\newcommand{\ionhy}{H{\sc ii}}
\newcommand{\kms}{kms$^{-1}$}

\newcommand{\lta}{\raisebox{-0.6ex}{$\,\stackrel
{\raisebox{-.2ex}{$\textstyle <$}}{\sim}\,$}}
\newcommand{\gta}{\raisebox{-0.6ex}{$\,\stackrel
{\raisebox{-.2ex}{$\textstyle >$}}{\sim}\,$}}   
\newcommand{\twospec}[2] {
 \begin{center}
    \begin{minipage}[t]{0.45\textwidth}
      \includegraphics[angle=270,scale=0.35]{#1}
    \end{minipage}
    \hfill
    \begin{minipage}[t]{0.45\textwidth}
      \includegraphics[angle=270,scale=0.35]{#2}
    \end{minipage}
  \end{center}
}
\newcommand{\onespec}[1] {
 \begin{center}
    \begin{minipage}[t]{0.45\textwidth}
      \includegraphics[angle=270,scale=0.35]{#1}
    \end{minipage}
  \end{center}	
}

\newcommand{\stackspec}[1] {
  \begin{center}
    \includegraphics[angle=270,scale=0.4]{#1}
  \end{center}
}

\shorttitle{37 and 38 GHz methanol masers}
\shortauthors{S. P. Ellingsen et al.}

\begin{document}

%\title{A search for 37 and 38 GHz class II methanol masers towards high-mass star formation regions}
\title{37 GHz methanol masers : Horsemen of the Apocalypse for the class II methanol maser phase?}

\author{S. P. Ellingsen$^\dagger$, S. L. Breen$^{\dagger\dagger}$}
\affil{School of Mathematics and Physics, University of Tasmania, 
  Private Bag 37, Hobart, TAS 7001, Australia}
\altaffiltext{$\dagger$}{Max Planck Institut f\"ur Radioastronomie, Auf dem H\"ugel 69, 53121 Bonn, Germany}
\altaffiltext{$\dagger\dagger$}{CSIRO Astronomy and Space Science, Australia Telescope National Facility, PO Box 76, Epping, NSW 1710, Australia}
\email{Simon.Ellingsen@utas.edu.au}
\author{A. M. Sobolev}
\affil{Ural Federal University, Lenin ave. 51, 620000 Ekaterinburg, Russia}
\author{M. A. Voronkov, J. L. Caswell}
\affil{CSIRO Astronomy and Space Science, Australia Telescope National Facility, PO Box 76, Epping, NSW 1710, Australia}
\author{N. Lo}
\affil{Departamento de Astronom\'ia, Universidad de Chile, Camino El Observatorio 1515, Las Condes, Santiago, Casilla 36-D, Chile}

\begin{abstract}
We report the results of a search for class II methanol masers at 37.7, 38.3 and  38.5~GHz towards a sample of 70 high-mass star formation regions.  We primarily searched towards regions known to show emission either from the 107 GHz class II methanol maser transition, or from the 6.035 GHz excited OH transition.  We detected maser emission from 13 sources in the 37.7 GHz transition, eight of these being new detections.  We detected maser emission from three sources  in the 38 GHz transitions, one of which is a new detection.  We find that 37.7~GHz methanol masers are only associated with the most luminous 6.7 and 12.2~GHz methanol maser sources, which in turn are hypothesised to be the oldest class II methanol sources.  We suggest that the 37.7 GHz methanol masers are associated with a brief evolutionary phase (of 1000-4000 years) prior to the cessation of class II methanol maser activity in the associated high-mass star formation region.
\end{abstract}


\section{Introduction}

Interstellar masers from a wide variety of molecular transitions are observed towards high-mass star formation regions.  In particular ground-state OH masers, 22 GHz water masers and 6.7 and 12.2 GHz methanol masers are commonly observed towards such regions.  Emission from one or more of these transitions has been observed towards more than 1000 sites within our Galaxy.  Methanol masers have been empirically divided into two classes.  The class I methanol masers are collisionally pumped and typically associated with weak shocks, such as where outflows interact with the surrounding ambient medium \citep[see][for a more detailed discussion]{Voronkov+06}.  The strongest and most common class I methanol masers are those at 36.2 and 44.1 GHz.  Class II methanol masers are radiatively pumped and often associated with OH and/or water masers and infrared sources.  The strongest and most common class II methanol masers are observed from the 6.7 and 12.2 GHz transitions.  This work focuses solely on class II masers. 

In addition to the strong and common transitions class II methanol maser transitions at 6.7 and 12.2 GHz there are a large number which are observed to exhibit maser emission in a smaller number of sources.  These sources typically show strong emission in the common transitions and a small number have been investigated in detail and the maser emission modelled \citep[e.g.][]{Cragg+01,Sutton+01}.  Maser pumping models are able to explain the broad properties of the observed maser transitions in that the strong and common transitions are usually strongly inverted over a wider range of physical conditions than the weaker, rarer transitions \citep{Cragg+02,Cragg+05,PK96a,PK96b}.  However, in many cases the models predict a wide range of parameter space for which emission from strong, common transitions would be accompanied by one or more rare, weak transitions.  If we assume that the models are broadly correct, then this suggests that the physical conditions in the majority of sources correspond to physical conditions which preclude the weak, rare masers.  This means that, in theory, by observing a wide variety of maser transitions towards a single source, the relative strength (or absence) of the different transitions can be used to constrain the physical conditions where the maser emission arises.  The sources where this technique has been applied in detail have been those which show emission in the largest number of rare OH and methanol transitions : W3(OH) \citep{Cesaroni+91,Sutton+01}, NGC6334F and G345.01+1.79 \citep{Cragg+01}.

Regions which show emission from an unusually large number of the rare, weak maser transitions (e.g. W3(OH), NGC6334F, G345.01+179), are by definition atypical.  They must either correspond to an intrinsically rare type of source, rare geometric orientation, or perhaps more likely, to a brief period during the evolution of high-mass star formation regions where the conditions are conducive for emission from a wide range of maser transitions.  To better understand the properties of more typical maser regions (those which exhibit emission in a smaller number of methanol maser transitions) we have undertaken a program to search for rare, weak masers towards a larger sample of sources, in particular those showing methanol maser emission in the 107 GHz transition.  We have previously searched this sample for the 19.9, 23.1, 85.5, 86.6 and 108.8 GHz class II methanol masers transitions \citep{Valtts+99,Ellingsen+03,Cragg+04,Ellingsen+04}.  Here we present the observations for a further three class II methanol transitions which are at frequencies near 38~GHz towards 70 high-mass star formation regions.  There is only one previous published study of these transitions, which detected maser emission towards 5 sources in the 37.7~GHz transition and towards two sources in the 38.3 and 38.5~GHz transitions \citep{HBM89}.  

\section{Observations} \label{sec:obs}

The primary sample in our study to expand our understanding of class II methanol masers in star formation regions has been the 25 sources observed to exhibit class II methanol maser emission in the 107~GHz methanol transition.  These sources were identified in searches towards more than 175 methanol maser sources \citep{Valtts+95,Valtts+99,Caswell+00,MB02}.  These are labelled as sample $A$ in Table~\ref{tab:nondet}.  Maser modelling, and previous observations suggests that the presence of a 107~GHz methanol maser in these sources makes them more likely to be associated with other rare, weak maser transitions than those sources without an associated 107~GHz maser.  With the exception of the three atypical sources previously identified, the majority of the sample are associated with either zero, or one rare, weak methanol maser transition, and so are an intermediate group between the majority of sources which are only observed in one or more of the 6.7 and 12.2 GHz transitions, and the very small number of sources which show maser emission in more than 6 different methanol transitions.  In order to reduce the potential for biases in our search we also included observations of two other groups of sources.  Previous 19.9~GHz methanol maser observations suggest that star formation regions with centimetre radio continuum emission and 6.035 GHz excited OH masers may be more likely to host rare, weak methanol masers \citep{Ellingsen+04}, and we observed all sources south of declination 0$^{\circ}$ meeting these criteria (sample $B$ in Table~\ref{tab:nondet}, 26 sources in total).   The final sample was a search towards all 6.7 GHz methanol masers with a peak flux density $> 75$ Jy, north of a declination of -15$^\circ$ in the catalogue of \cite{Malyshev+03} (sample $C$ in Table~\ref{tab:nondet}, 17 sources).  Due to time constraints G$\,30.70-0.07$ and G$\,31.28+0.06$ (nominally members of sample $C$ were not observed), while ON1 and DR21(OH) were observed despite not meeting the flux density criteria due to the small number of target sources in that LST range. The total sample comprises 70 sources, the vast majority (67) of which are known 6.7 GHz methanol maser sources.  The exceptions are Orion KL, OMC 2 (observed as system test sources at Onsala) and G328.307+0.430 (a 6.035 GHz OH maser source without an associated 6.7 GHz methanol maser).  For some of the sources more accurate positions have recently been published \citep[e.g.][]{Caswell09} which show some of the observed positions were in error by up to 10\arcsec.  These offsets are of comparable magnitude to the telescope pointing errors and have no measurable effect on the observations, however, it should be noted that the positions listed in Table~1, are in all cases the positions at which the observations were made and in some cases these are no longer the most accurate positions measured for these sources.

In order to search for 37.7 and 38 GHz methanol masers in both the northern and southern hemispheres, two separate observing programs were undertaken.  The southern hemisphere observations were carried out with the Australia Telescope National Facility (ATNF) Mopra 22m radio telescope during 2009 May 30 - June 3 and 2011 February 1 \& 3.  The observations were made with the 7mm receiver system as a back-up project for periods when the weather was too poor for observations at 3mm.  The system temperature varied between approximately 70 and 240 K during the observations.  The Mopra spectrometer (MOPS) was configured with 14 IF bands (``zooms'') spread over the frequency range from 33.1 -- 40.0 GHz.  Each IF band covered 138 MHz, with 4096 spectral channels per band and two orthogonal linear polarizations were recorded.  This configuration yields a velocity coverage of approximately 1000 \kms\/ and velocity resolution of 0.32~\kms\/ (for a channel spacing of 0.27~\kms) for unsmoothed spectra.  The Mopra telescope has RMS pointing errors of $<$ 10 $\arcsec$ and at 38~GHz the telescope has a half-power beam-width of 73 $\arcsec$ \citep{Urquhart+10}.

The 14 IF bands were arranged to cover a variety of methanol maser and thermal molecular transitions.  The four methanol maser transitions observed were the class~I $4_{-1} \rightarrow 3_{0}E$ transition at 36.2 GHz and the $7_{-2} \rightarrow 8_{-1}E$, $6_{2} \rightarrow 5_{3}\mbox{A}^-$ and $6_{2} \rightarrow 5_{3}\mbox{A}^+$ class II transitions at 37.7, 38.3 and 38.5 GHz, respectively.  The thermal lines observed, included a variety of cyanopolyyne transitions and CH$_3$CN ($J=2-1$).  With the exception of the HC$_{3}$N ($J=4-3$) ground-state transition at 36.392 GHz, most of these thermal transitions were not detected, or were only detected towards a small number of sources with low signal to noise.  In this paper we discuss only the results of the observations of the class~II methanol maser transitions, the results for the 36.2~GHz class I maser line and the thermal emission will be reported in a future publication.

The observations were made as a series of position-switched integrations of approximately 60 seconds duration, with reference observations offset from the target position by 5 $\arcmin$ in declination.  The data were processed using the ASAP (ATNF Spectral Analysis Package).  Most sources had an on-source integration time of approximately 230 seconds, with the alignment of the velocity channels for the individual scans carried out during processing.  The system temperature for the observations was measured by a continuously switched noise diode.  At a frequency of 38~GHz the Mopra telescope has a main beam efficiency of approximately 0.52, which implies a scaling factor of 14 for conversion of the intensity scale from units of antenna temperature to Janskys \citep{Urquhart+10}. The RMS noise level in the final spectra (after averaging over polarizations and time, but with no smoothing) varied between 0.4 -- 1.6 Jy, with the majority being $<$0.7 Jy.  The weather data for Mopra during the observing period are not available and we have used the weather observations at the ATCA (approximately 100 km away) to estimate the atmospheric opacity. The estimated zenith opacity during the observations varied between 0.06 and 0.09, which implies attenuation of between 6 and 11\% for these observations.  Taking into account pointing, flux density calibration and variations in the opacity we estimate the measured flux densities to be accurate to 15\%.

Two sources (G$\,351.581-0.353$ and G$\,336.018-0.827$) were accidentally omitted from the 2009 Mopra observations and these were observed using a director's time allocation on 2011 February 1 or 3.  We also made an additional observation of G$\,35.201-1.736$ on 2011 February 3.

The northern observations were carried out using the Onsala 20m telescope during 2005 November 29 - December 6 using a HEMT receiver operating in the 36.0 - 49.8 GHz range.  The system temperature varied between approximately 150 and 350 K during the observations.  The spectrometer was configured with 12.8 MHz bandwidth and 1600 spectral channels for the single polarization recorded.  This configuration yields a velocity coverage of approximately 100 \kms\/ and velocity resolution of approximately 0.06~\kms\/  for unsmoothed spectra.  The observations were made in dual beam switching mode with a frequency of 2 Hz and with reference observations offset from the target position by 11$\arcmin$ in declination.  The data were processed using the XS package written by P. Bergman (http://www.chalmers.se/rss/oso-en/observations/data-reduction-software).  The on-source integration time varied between 15 and 160 minutes.  The system temperature for the observations was measured using the chopper-wheel method.  At a frequency of 38~GHz the Onsala 20~m telescope has a FWHM of $100 \arcsec$ and the aperture efficiency is approximately 0.53, which implies a scaling factor of 18 for conversion of the intensity scale from units of antenna temperature to Janskys. The RMS noise level in the final spectra was typically in the range 0.5 -- 2.2 Jy.

% Rest frequency calculations:
% The corrections are the velocities to _add_ to the observed velocities from Haystack/Onsala to put them on the same rest frequency scale as Mopra.
% Mopra		Haystack		Onsala		Haystack corr (km/s)	Onsala corr (km/s)
%37.703696	37.703729	37.703700	-0.26				-0.03
%38.293292	38.293306	38.293268	-0.11				+0.19
%38.452652	38.452662	38.452677	-0.08				-0.20
% These velocity corrections have been applied to the Onsala spectra.
% NOTE: These are the same rest frequencies used by Sutton et al. 2001 in their W3(OH modeling).
% The Onsala values are those from Muller, Menten & Mader 2004 and have uncertainties of 30, 50 and 50 kHz respectively.
% At 37.7 GHz, 30 kHz corresponds to a velocity of 0.24 km/s and 50 kHz corresponds to 0.4 km/s

We adopted rest frequencies of 37.703696, 38.293292 and 38.452652 GHz for the $7_{-2} \rightarrow 8_{-1}E$, $6_{2} \rightarrow 5_{3}A^-$ and $6_{2} \rightarrow 5_{3}A^+$ transitions respectively \citep{Xu+97}.  Comparing these to the rest frequencies used by \citet{HBM89}, the values we have adopted are between 10 and 33 kHz lower in frequency.  To directly compare the velocities observed by \citet{HBM89} with those we have observed, offsets of -0.26, -0.11 and -0.08~\kms\/ should be added to their values for the velocities of the 37.7, 38.3 and 38.5 GHz emission respectively.  Where we have made such comparisons throughout the paper we have applied these offsets to the values reported for the \citet{HBM89} velocities.  The uncertainties in the rest frequencies of \citet{Xu+97} are around 50~kHz \citep{Muller+04}, which corresponds to a velocity of 0.4~\kms\/ at 37.7~GHz.  The close agreement between the velocity of the 37.7~GHz masers and peaks in other transition (see section~\ref{sec:results}) suggests that the adopted frequencies are more accurate than the formal uncertainties for these transitions.

\section{Results} \label{sec:results}

We have used the Mopra and Onsala telescopes to undertake a search for emission in the 37.7 GHz class~II methanol transition ($7_{-2} \rightarrow 8_{-1}E$) towards a total of 70 sources.  Forty six of these sources were observed with the Australia Telescope National Facility Mopra 22~m telescope, and for these simultaneous observations of both the $6_{2} \rightarrow 5_{3}A^-$ and $6_{2} \rightarrow 5_{3}A^+$ (38.3 and 38.5 GHz respectively) were also obtained.  Of the 29 sources observed with the Onsala telescope at 37.7 GHz, separate observations of the 38.3~GHz transition were made towards 27, and of the 38.5 GHz transition towards 4.  The position searched, RMS noise level and velocity range covered for the observed sources for all three transitions are given in Table~\ref{tab:nondet}.

% NOTE: Observations were made of the following sources, but there are problems:
% G351.581: In the initial observations in 2009 the wrong coordinates were used - 
%                     the RA from G351.581 with the declination from 351.775.
%                     Data listed is from a reobservation in 2011 Feb 1.
% W51 : The position observed at Onsala was incorrect.
% ON2 : The position observed at Onsala was off by 1.3 arcminutes - more than the primary beam.
% 81.87+0.78 (W75N) : The position is offset by 23 arcsec in declination, but that isn't off enough
%                                        to be worried about.

%The information on sources searched etc in the results is based on the numbers in the table below.
% If this is modified change the numbers there as well.
\begin{table*}
\begin{center}
\caption{Sources observed in one or more of the 37.7, 38.3 or 38.5 GHz class II methanol maser transitions.  No entry in the column means that the source was not observed in that transition (this only applies to the Onsala observations as all transitions were observed simultaneously with Mopra).  An asterisk for a transition means that emission was detected in that transition, the reported RMS was measured in a line-free region of the spectrum.  Further details for the detected sources are contained in Tables~\ref{tab:results37} -- \ref{tab:results385}.  Mopra observations marked with a $^\dagger$ were undertaken on either 2011 February 1 or 3.} \label{tab:nondet}
\begin{tabular}{lccrrrccc}
\tableline
\multicolumn{1}{c}{\bf Source} & \multicolumn{1}{c}{\bf Right Ascension} & \multicolumn{1}{c}{\bf Declination} & \multicolumn{3}{c}{\bf RMS (Jy)} & \multicolumn{1}{c}{\bf Velocity Range} & \multicolumn{1}{c}{\bf Sample} & \multicolumn{1}{c}{\bf Telescope}  \\
\multicolumn{1}{c}{\bf Name} & \multicolumn{1}{c}{\bf (J2000)} & \multicolumn{1}{c}{\bf (J2000)} & \multicolumn{1}{c}{\bf 37.7 GHz} &  \multicolumn{1}{c}{\bf 38.3 GHz} & \multicolumn{1}{c}{\bf 38.5 GHz} &\multicolumn{1}{c}{\bf (\kms)} & & \\
\tableline
% All RMS for Mopra observations have been corrected for updated sensitivity
W3(OH)                      & 02:27:03.8 & +61:52:25 & 0.3* & 0.5* & 1.1* &  -93 -- 7    & A & Onsala \\ 
Orion KL                     & 05:35:14.5 & -05:22:30 & 0.4* &  0.6* & 1.2*  &  -42 -- 58   &  & Onsala \\
OMC\,2                      & 05:35:27.5 & -05:09:37 & 0.7 &  1.0  &       &   -38 -- 62    &  & Onsala \\
Orion S6                     & 05:35:14.0 & -05:24:05 & 0.7 &  0.8  &       &  -51 -- 49    & C & Onsala \\
G$\,173.481+2.446$ & 05:39:12.9 & +35:45:54 & 0.6 &  0.6  &       &  -64 -- 36  & C & Onsala \\
G$\,188.95+0.89$     & 06:08:53.7 & +21:38:30 & 0.8* &  0.6  & 1.1 &  -39 -- 61  & A & Onsala \\
G$\,192.600-0.048$  & 06:12:54.5 & +17:59:20 & 0.8 & 0.7   &       &  -46 -- 54   & A & Onsala \\
G$\,213.71-12.60$    & 06:07:48.0 & -06:22:57 & 0.9 &  0.7  &       &   -38 -- 62    & C & Onsala \\
G$\,240.316+0.071$ & 07:44:51.9 & -24:07:42 & 1.4 & 1.3 & 1.3 & -40 -- 160   & B & Mopra \\
G$\,300.969+1.148$ & 12:34:53.4 & -61:39:40 & 1.9 & 1.6 & 1.5 & -140 -- 60   & B & Mopra \\
G$\,309.921+0.479$ & 13:50:41.8 & -61:35:10 & 1.6 & 1.5 & 1.5 & -160 -- 40   & B & Mopra \\
G$\,310.144+0.760$ & 13:51:58.5 & -61:15:40 & 0.4 & 0.4 & 0.4 & -200 -- 100 & A & Mopra \\
G$\,311.643-0.380$  & 14:06:38.8 & -61:58:23 & 1.4 & 1.5 & 1.5 & -70 -- 130   & B & Mopra \\
G$\,318.948-0.196$  & 15:00:55.4 & -58:58:53 &  0.4* & 0.5 & 0.4 & -180 -- 120 & A & Mopra \\
G$\,323.459-0.079$  & 15:29:19.4 & -56:31:20 & 1.6 & 1.5 & 1.6 & -170 -- 30   & B & Mopra \\
G$\,323.740-0.263$  & 15:31:45.6 & -56:30:50 &  0.8* & 0.4 & 0.4 & -200 -- 100 & A & Mopra \\
G$\,327.120+0.511$ & 15:47:32.8 & -53:52:38 & 0.4 & 0.4 & 0.4 & -240 -- 60   & A & Mopra \\
G$\,328.237-0.547$  & 15:57:58.3 & -53:59:22 & 1.8 & 1.5 & 1.5 & -150 -- 50   & B & Mopra \\
G$\,328.307+0.430$ & 15:54:06.5 & -53:11:40 & 1.4 & 1.4 & 1.4 & -190 -- 10   & B & Mopra \\
G$\,328.808+0.633$ & 15:55:48.5 & -52:43:07 & 0.5 & 0.5 & 0.5 & -200 -- 100 & A & Mopra \\
G$\,330.953-0.182$  & 16:09:52.8 &  -51:54:56 & 1.1 & 1.2 & 1.1 & -190 -- 10   & B & Mopra \\
G$\,331.542-0.066$  & 16:12:09.1 & -51:25:48 & 0.8 & 0.8 & 0.8 & -190 -- 10   & B & Mopra \\
G$\,336.018-0.827$  & 16:35:09.3 & -48:46:47 & 0.6 & 0.6 & 0.6 & -200 -- 100 & A & Mopra$^\dagger$ \\
G$\,337.705-0.053$  & 16:38:29.7 & -47:00:35 &  0.6* & 0.6 & 0.5 & -205 -- 95   & B & Mopra \\
G$\,337.404-0.402$  & 16:38:50.4 & -47:28:03 & 0.5 & 0.6 & 0.6 & -140 -- 60   & B & Mopra \\
G$\,338.075+0.012$ & 16:39:39.0 & -46:41:28 & 0.6 & 0.5 & 0.5 & -205 -- 95   & B & Mopra \\
G$\,340.054-0.244$  & 16:48:13.9 & -45:21:44 & 0.6 & 0.7 & 0.6 & -200 -- 100 & A & Mopra \\
G$\,340.785-0.096$  & 16:50:14.8 & -44:42:25 & 0.6* & 0.7 & 0.6 & -250 -- 50   & A & Mopra \\
G$\,339.884-1.259$  & 16:52:04.8 & -46:08:34 & 0.8* & 0.8 & 0.8 & -190 -- 110 & A & Mopra \\
G$\,345.010+1.792$  & 16:56:47.7 & -40:14:26 & 0.7* & 0.7* & 0.6* & -170 -- 130 & A & Mopra \\
G$\,343.929+0.125$ & 17:00:10.9 & -42:07:19 & 0.5 & 0.5 & 0.5 & -190 -- 110 & B & Mopra \\
G$\,345.504+0.348$ & 17:04:22.8 & -40:44:21 & 0.7 & 0.7 & 0.7 & -170 -- 130 & A & Mopra \\
G$\,345.003-0.223$  & 17:05:10.9 & -41:29:06 & 0.7 & 0.7 & 0.7 & -170 -- 130 & A & Mopra \\
G$\,347.628+0.148$ & 17:11:50.9 & -39:09:30 & 0.6 & 0.5 & 0.5 & -250 -- 50   & B & Mopra \\
G$\,348.550-0.979$  & 17:19:20.4 & -39:03:52 & 0.7 & 0.6 & 0.6 & -110 -- 90   & B & Mopra \\
G$\,348.703-1.043$  & 17:20:04.1 & -38:58:30 & 0.8* & 0.7 & 0.7 & -150 -- 150 & A & Mopra \\
NGC6334F                 & 17:20:53.4 & -35:47:00 &  0.9* & 1.0* & 1.0* & -150 -- 150 & A & Mopra \\
G$\,351.775-0.536$  & 17:26:42.7 & -36:09:16 & 0.7 & 0.6 & 0.6 & -160 -- 140 & B & Mopra \\
G$\,351.581-0.353$  & 17:25:25.2 & -36:12:46 & 0.5 & 0.5 & 0.5 & -250 -- 50   & B & Mopra$^\dagger$ \\
G$\,353.410-0.360$  & 17:30:26.2 & -34:41:45 & 0.6 & 0.6 & 0.6 & -170 -- 130 & A & Mopra \\
G$\,354.724+0.300$ & 17:31:15.5 & -33:14:05 & 1.6 & 1.6 & 1.5 & -50 -- 250   & B & Mopra \\
G$\,355.344+0.147$ & 17:33:29.0 & -32:47:59 & 1.3 & 1.4 & 1.3 & -130 -- 170 & B & Mopra \\
G$\,3.910+0.001$      & 17:54:38.8 & -25:34:42 & 1.1 & 0.8 & 0.7 & -130 -- 170 & B & Mopra \\
G$\,5.885-0.392$       & 18:00:30.4 & -24:04:03 & 0.9 & 1.0 & 0.9 & -150 -- 150 & B & Mopra \\
G$\,8.669-0.356$       & 18:06:19.0 & -21:37:32 & 1.3 & 1.2 & 1.1 & -100 -- 200 & B & Mopra \\
G$\,9.621+0.196$      & 18:06:14.8 & -20:31:32 &  0.6* & 0.6 & 0.6 & -150 -- 150 & A & Mopra \\
                                       & 18:06:14.8 & -20:31:40 &  1.7* &        &       &   -50 -- 50    &    & Onsala \\
G$\,10.623-0.384$     & 18:10:28.7 & -19:55:49 & 1.2 & 1.2 & 1.1 & -150 -- 150 & B & Mopra \\
G$\,11.904-0.141$     & 18:12:11.5 & -18:41:28 & 1.2 & 1.4 & 1.2 & -110 -- 190 & B & Mopra \\
G$\,12.909-0.260$     & 18:14:39.5 & -17:52:00 & 0.4 & 0.4 & 0.4 & -110 -- 190 & A & Mopra \\
G$\,15.034-0.677$     & 18:20:24.8 & -16:11:34 & 1.1 & 1.1 & 1.2 & -130 -- 170 & B & Mopra \\
G$\,20.23+0.07$         & 18:27:43.9 & -11:15:04 & 1.1 & 1.2 &       & 22 -- 122    & C & Onsala \\
G$\,23.440-0.182$     & 18:34:39.2 & -08:31:24 &  0.4* & 0.4 & 0.4 & 0 -- 200       & A & Mopra \\
                                       & 18:34:39.0 & -08:31:36 & 1.0 &  1.3 &       & 47 -- 147    &   & Onsala \\
G$\,23.010-0.411$     & 18:34:40.4 & -09:00:37 & 0.3 & 0.3 & 0.3 & -80 -- 220   & A & Mopra \\
                                       & 18:34:39.9 & -09:00:44 & 1.0 & 0.9 &        & 26 -- 126    &   & Onsala \\
G$\,25.65+1.05$         & 18:34:19.7 & -05:59:44 & 1.0 & 1.6 &       & -8 -- 92       & C & Onsala \\
G$\,25.709+0.044$    & 18:38:03.1 & -06:24:15 &  1.2 &  1.0 &       & 45 -- 145    & C & Onsala \\
G$\,28.201-0.049$     & 18:42:58.2 & -04:13:56 & 1.5 & 1.4 & 1.4 & -50 -- 250   & B & Mopra \\
                                       & 18:42:58.0 & -04:14:01 & 4.9 &        &       &   49 -- 149   & & Onsala \\
G$\,29.950-0.020$     & 18:46:03.6 & -02:39:24 & 0.8 & 1.1  &       & 46 -- 146    & C & Onsala \\
G$\,32.03+0.05$         & 18:49:38.5 & -00:45:29 & 2.2 & 1.7  &       & 43 -- 143    & C & Onsala \\
G$\,33.640-0.210$     & 18:53:28.7 & +00:31:58 & 2.2 &         &       & 10 -- 110    & C & Onsala \\
G$\,37.43+1.50$         & 18:54:17.2 & +04:41:09 & 2.2 & 1.8 &        & -9 -- 91       & C & Onsala \\
G$\,35.20-0.74$          & 18:58:13.0 & +01:40:33 & 0.9 & 1.2  &       & -22 -- 78     & C & Onsala \\
G$\,35.201-1.736$     & 19:01:45.5 & +01:13:36 & 0.5 & 0.5 & 0.5 & -110 -- 190 & A & Mopra \\
                                       & 19:01:45.5 & +01:13:29 &  0.8* & 0.9 & 1.4 & -7 -- 93       &  & Onsala \\
G$\,43.80-0.13$          & 19:11:53.8 & +09:35:46 & 1.7 &       &        & -10 -- 90    & C & Onsala \\
G$\,49.49-0.39$          & 19:23:43.4 & +14:30:35 & 0.8 & 0.8 &        & 9 -- 109     & C & Onsala \\
ON1                               & 20:10:09.2 & +31:31:35 & 0.8 & 0.8 &        & -35 -- 65     & C & Onsala \\
DR21(OH)                    & 20:39:00.7 & +42:22:51 & 0.6 &  0.7  &       &  -50 -- 50   & C & Onsala \\
G$\,81.87+0.78$          & 20:38:36.8 & +42:38:00 & 0.5 &  0.7  &       &  -43 -- 57   & C & Onsala \\
G$\,108.18+5.52$        & 22:28:52.1 & +64:13:43 & 0.6 &  0.6  &       &  -61 -- 38   & C & Onsala \\
Cep A                             & 22:56:18.1 & +62:01:49 & 0.7 &  0.5  &       &  -52 -- 47   & A & Onsala \\
NGC7538                     & 23:13:45.3 & +61:28:11 & 0.7 &  0.6  &       &  -106 -- - 6   & A & Onsala \\
\tableline
\end{tabular}
\end{center}
\end{table*}

Maser emission from the 37.7~GHz $7_{-2} \rightarrow 8_{-1}E$ transition was detected towards 13 sources.  Quasi-thermal 37.7 GHz emission was detected towards Orion KL, which was observed as part of system testing.  There are several reasons why we are confident that (with the exception of Orion KL) we are observing maser emission in these transitions, despite the peak flux density of some of the sources being $<$ 5 Jy.  For about half the sources, the large peak flux density and narrow line width ($<$ 0.5 \kms\/ in most cases), leaves little doubt that this is maser emission.  For the remaining sources the emission is also narrow and coincides in velocity with strong spectral features in other methanol maser transitions (see Section~\ref{sec:indiv}). The spectra of the detected maser sources are shown in Fig.~\ref{fig:masers37} and the properties of the emission are summarized in Table~\ref{tab:results37}.  Combining the results of the 37.7~GHz maser search of \citet{HBM89} with those reported here, a total of 77 class~II methanol maser sites have been searched for the 37.7~GHz transition, resulting in a total of 13 detections of masers.  Eight of the detections (all in the southern hemisphere) are reported here for the first time and this increases the number of sources which are known to show maser emission in this transition by more than a factor of 2.

\begin{figure}
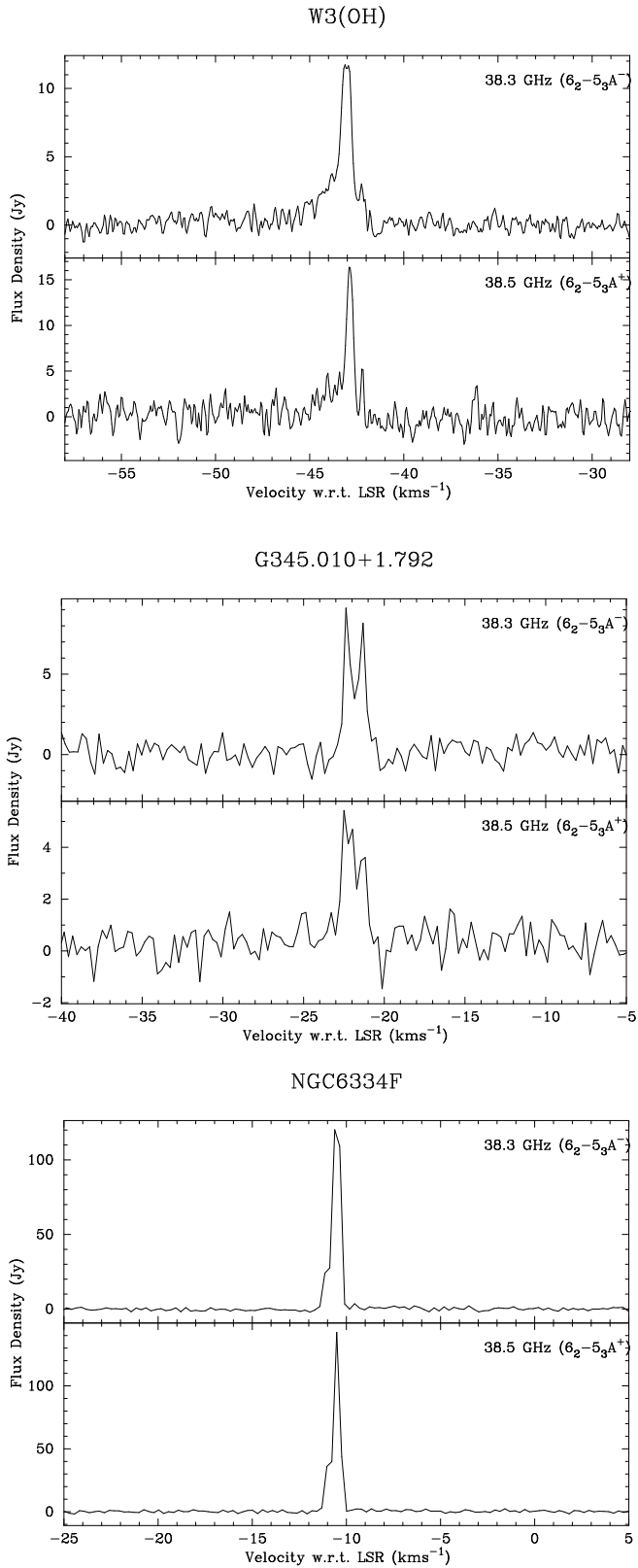

\stackspec{w3oh_38ghz.eps}
\stackspec{G345.010_38ghz.eps}
\stackspec{NGC6334F_38ghz.eps}
\caption{Spectra of masers detected in the 38.3 and 38.5 GHz transitions.} \label{fig:masers38}
\end{figure}

Maser emission from the $6_{2} \rightarrow 5_{3}A^-$ and $6_{2} \rightarrow 5_{3}A^+$ transitions (38.3 and 38.5 GHz respectively) was detected towards 3 sources,  one of which (G$\,345.010+1.792$) is a new detection.  The spectra of the 38.3 and 38.5 GHz masers are shown in Fig.~\ref{fig:masers38} and the properties of the emission are summarized in Tables~\ref{tab:results383} \& \ref{tab:results385}.  

\begin{figure*}
\twospec{w3oh_37ghz.eps}{G188.95_37ghz.eps}
\twospec{G318.948_37ghz.eps}{G323.740_37ghz.eps}
\twospec{G337.705_37ghz.eps}{G339.884_37ghz.eps}
\twospec{G340.785_37ghz.eps}{G345.010_37ghz.eps}
\caption{Spectra of masers detected in the 37.7 GHz transition.} \label{fig:masers37}
\end{figure*}

\begin{figure*}
\twospec{G348.703_37ghz.eps}{NGC6334F_37ghz.eps}
\twospec{G9.621_37ghz.eps}{G23.440_37ghz.eps}
\onespec{G35.201_37ghz.eps}
\caption{Spectra of masers detected in the 37.7 GHz transition {\em continued...}.} \label{fig:masers37_cont}
\end{figure*}

% All fitting redone with ASAP3.0 post correction of the Mopra flux density scale.
% Parameters from Onsala fits (except the peak intensity) taken from Andrej's email of 3 December 2010, velocity scale modified by -0.03 km/s for W3(OH) and G35.201
% Parameters for peak intensity from Andrej's email of 26 August 2011 which states "I made analysis of the uncertainties of the peak flux densities for the Onsala observations assuming that the relative error of the peak flux density is equal to the relative error of the quotient of the integrated antenna temperature and the line width and using the error propagation formula"

\begin{table*}
\begin{center}
\caption{Characteristics of the sources detected in the 37.7 GHz methanol transition.  For parameters where the error is indicated by a dash that particular parameter was held fixed in the Gaussian fit.} \label{tab:results37}
\begin{tabular}{cccrrr}
\tableline
\multicolumn{1}{c}{\bf Source} & \multicolumn{1}{c}{\bf Right Ascension} & \multicolumn{1}{c}{\bf Declination} & \multicolumn{1}{c}{\bf Peak Flux} & \multicolumn{1}{c}{\bf Velocity} & \multicolumn{1}{c}{\bf Full width half} \\
\multicolumn{1}{c}{\bf Name} & \multicolumn{1}{c}{\bf (J2000)} & \multicolumn{1}{c}{\bf (J2000)} & \multicolumn{1}{c}{\bf Density(Jy)} & \multicolumn{1}{c}{\bf (\kms)} & \multicolumn{1}{c}{\bf maximum (\kms)} \\
\tableline
W3(OH)                       & 02:27:03.8 & +61:52:25 & 2.2(0.2)     & -42.96(0.03) & 1.01(0.05) \\
G$\,188.95+0.89$     & 06:08:53.7 & +21:38:30 & 23.4(0.6)   & 10.74(0.01) & 0.68(0.01) \\
G$\,318.948-0.196$ & 15:00:55.4 & -58:58:53 & 9.3(0.5)   & -34.19(0.01) & 0.47(0.03) \\
G$\,323.740-0.263$ & 15:31:45.6 & -56:30:50 & 14.6(0.5) &-51.18(0.01) & 0.64 (0.02) \\
                                   &                      &                      &  3.7(0.4)  &-49.7(0.05)     & 1.04 (0.12) \\
G$\,337.705-0.053$ & 16:38:29.7 & -47:00:35 & 2.4(0.6)  & -55.00(0.06) & 0.46 (0.14) \\
G$\,339.884-1.259$ & 16:52:04.8 & -46:08:34 & 323(0.9) & -38.70(0.001) & 0.38 (0.001) \\
G$\,340.785-0.096$ & 16:50:14.8 & -44:42:25 & 20.5(0.7) & -105.48 (0.01) & 0.38 (0.01)  \\
                                     &                      &                   & 12.4(0.7) & -106.57 (0.01) & 0.39 (0.03)  \\
G$\,345.010-1.792$ & 16:56:47.7 & -40:14:26 & 207(1.0) & -22.10(-) & 0.33(0.001) \\
                                     &                      &                   & 93.8(0.7) & -21.83(-) & 0.72(0.006) \\
                                     &                      &                   & 11.0(1.1) & -20.90(-) & 0.26(0.03) \\
G$\,348.703-1.043$ & 17:20:04.1  & -38:58:30 & 4.4(0.9)   & -3.40(-) & 0.36(0.08) \\
NGC6334F                & 17:20:53.4 & -35:47:00 & 70.4(1.5) & -10.91(0.01) & 0.36(0.01) \\
                                     &                     &                    & 21.9(1.2) & -10.45(0.02) & 0.4(-) \\
                                     &                     &                    & 3.5(0.6)   & -10.1(0.2)   & 2.9(0.4) \\
G$\,9.621+0.196$    & 18:06:14.8 & -20:31:32 & 23.6(0.6) & -1.08(0.01) & 0.59(0.01) \\
                                     &                     &                    & 8.3(0.7)  & -0.28(0.02) & 0.43(0.05) \\
G$\,23.440-0.182$  & 18:34:39.2 & -08:31:24  & 2.3(0.3)  & 98.03(0.04) & 0.5(-) \\
G$\,35.201-1.736$ & 19:01:45.5 & +01:13:29 & 3.2(1.3)    & 44.51(0.03) & 0.53(0.2) \\
\tableline
\end{tabular}
\end{center}
\end{table*}

% Parameters from Onsala fits taken from Andrej's email of 3 December, velocity scale modified by +0.19 km/s for W3(OH).  See comment on intensity uncertainty in 37 GHz table.
\begin{table*}
\begin{center}
\caption{Characteristics of the sources detected in the 38.3 GHz methanol transition.  For parameters where the error is indicated by a dash that particular parameter was held fixed in the Gaussian fit.} \label{tab:results383}
\begin{tabular}{cccrrr}
\tableline
\multicolumn{1}{c}{\bf Source} & \multicolumn{1}{c}{\bf Right Ascension} & \multicolumn{1}{c}{\bf Declination} & \multicolumn{1}{c}{\bf Peak Flux} & \multicolumn{1}{c}{\bf Velocity} & \multicolumn{1}{c}{\bf Full width half} \\
\multicolumn{1}{c}{\bf Name} & \multicolumn{1}{c}{\bf (J2000)} & \multicolumn{1}{c}{\bf (J2000)} & \multicolumn{1}{c}{\bf Density(Jy)} & \multicolumn{1}{c}{\bf (\kms)} & \multicolumn{1}{c}{\bf maximum (\kms)} \\
\tableline
W3(OH)                       & 02:27:03.8 & +61:52:25 & 10.9(0.4)     & -43.09(0.008) & 0.91(0.03) \\
G$\,345.010-1.792$ & 16:56:47.7 & -40:14:26 & 9.4(0.7) & -22.30(0.02) & 0.46(0.05) \\
                                     &                      &                   & 7.6(0.6) & -21.38(0.03) & 0.61(0.07) \\
NGC6334F                & 17:20:53.4 & -35:47:00 & 174(7.0) & -10.49(0.001) & 0.34(0.02) \\
                                     &                     &                    & 32.9(2.7) & -11.02(0.008) & 0.38(0.04) \\
\tableline
\end{tabular}
\end{center}
\end{table*}

% Parameters from Onsala fits taken from Andrej's email of 3 December, velocity scale modified by -0.2 km/s for W3(OH). See comment on intensity uncertainty in 37 GHz table.
\begin{table*}
\begin{center}
\caption{Characteristics of the sources detected in the 38.5 GHz methanol transition.  For parameters where the error is indicated by a dash that particular parameter was held fixed in the Gaussian fit.} \label{tab:results385}
\begin{tabular}{cccrrr}
\tableline
\multicolumn{1}{c}{\bf Source} & \multicolumn{1}{c}{\bf Right Ascension} & \multicolumn{1}{c}{\bf Declination} & \multicolumn{1}{c}{\bf Peak Flux} & \multicolumn{1}{c}{\bf Velocity} & \multicolumn{1}{c}{\bf Full width half} \\
\multicolumn{1}{c}{\bf Name} & \multicolumn{1}{c}{\bf (J2000)} & \multicolumn{1}{c}{\bf (J2000)} & \multicolumn{1}{c}{\bf Density(Jy)} & \multicolumn{1}{c}{\bf (\kms)} & \multicolumn{1}{c}{\bf maximum (\kms)} \\
\tableline
W3(OH)                       & 02:27:03.8 & +61:52:25 & 16.1(1.1)     & -42.88(0.009) & 0.47(0.03) \\
G$\,345.010-1.792$ & 16:56:47.7 & -40:14:26 & 5.0(0.5) & -22.25(0.06) & 1.04(0.15) \\
                                     &                      &                   & 3.6(0.8) & -21.26(0.05) & 0.48(0.13) \\
NGC6334F                & 17:20:53.4 & -35:47:00 & 150(20) & -10.47(0.06) & 0.33(0.11) \\
                                     &                     &                    & 44(15) & -10.94(0.11) & 0.37(0.09) \\
\tableline
\end{tabular}
\end{center}
\end{table*}

\subsection{Comments on individual sources of interest} \label{sec:indiv}

% 38 GHz peaks (and 107) are towards red-shifted end of spectrum.  6.7 GHz velocity range is -48 -- -41 km/s.
{\bf W3(OH) (G$\,\mathbf{133.94+1.04}$) :} This is the archetypal class II methanol and OH maser source.  It is one of the 5 sources with detected maser emission at 37.7~GHz in the only previous publication studying this transition \citep{HBM89}.  In a detailed study of class II methanol masers in W3(OH) \citet{Sutton+01} identified the 37.7 GHz emission as having an unusual profile compared to the other class II transitions and requested careful reobservation to confirm its nature. Our Onsala observations show a very similar spectrum to that observed by \citet{HBM89}, although the peak flux density peak is somewhat weaker the velocities agree (once account is taken of the different rest frequencies).  W3(OH) also shows emission from both the 38.3 and 38.5~GHz transitions.  The peak intensity of the emission detected in the Onsala observations is similar to that observed by \citet{HBM89}, as is the velocity of the narrow emission (we have not attempted to fit the ``pedestal'' component).  The spectrum of these two transitions is similar to that seen in the $7_{2} \rightarrow 6_{3}\mbox{A}^-$ and $7_{2} \rightarrow 6_{3}\mbox{A}^+$  class II transitions at 86.6 and 86.9 GHz \citep{Sutton+01}, which are in the same transition series.  The 107 GHz masers also show a strong narrow peak at a velocity of -43.1~\kms\/.  So in summary, the relatively weak maser emission first detected in these three transitions by \cite{HBM89} has changed little over the approximately 17 year interval between those observations and the ones reported here.

% No 19.9 GHz detection > 0.2 Jy ; 6.7 GHz peak of 495 Jy at 11 km/s (range -4 -- +12 km/s) ; 12.2 GHz peak 235 Jy at 10.4 km/s, 107 GHz 15.5 Jy at 10.9 km/s ; 37.7 GHz peak of 23.4 Jy at 10.7 km/s.  NOTE: The emission here is at the far red-shifted end of the spectrum.  HBM89 report ~ 12 Jy at a velocity of 10.8 km/s (with their corrected rest frequency).
{\bf G$\,\mathbf{188.95+0.89}$ :} This source hosts a moderately strong 6.7 GHz methanol maser with a peak flux density of approximately 500~Jy at a velocity of 11~\kms\/ \citep{Caswell+95a}.  All of the other class II masers observed in this source also have their peak emission at velocities of around 10-11~\kms.  The 37.7~GHz emission in this source was first detected by \citet{HBM89}, who observed it to have a peak flux density of approximately 12~Jy.  Our Onsala observations detect emission at the same velocity, with a peak flux density approximately a factor of two greater.

% No 19.9 GHz detection $>$ 0.17 Jy ; 6.7 GHz peak of 780 Jy at -35 kms (about 580 Jy at -34.6 km/s in MMB) ; 6.7 GHz range -39 - -31 km/s ; 12.2 GHz 180 Jy at -34.5 km/s ; 107 GHz peak 5.7 Jy at -34.2 km/s
{\bf G$\,\mathbf{318.948-0.196}$ :} This is a moderate intensity 6.7 GHz methanol maser for which the peak emission in all the detected class II methanol maser transitions (6.7, 12.2, 37.7 and 107 GHz) is at around -34.5~\kms\/ \citep{Caswell+95a,Caswell+95b,Caswell+00}, approximately the middle of the total velocity range of the 6.7 and 12.2 GHz emission.
 
 % 37.7 GHz couldn't fit a third Gaussian to the 2 Jy shoulder at -51.8 km/s.19.9 GHz, 0.15 Jy, -51.4 km/s, FWHM 3.1 km/s
 % 6.7 GHz peak of 2860 Jy at -51 km/s, range -54 -- -47 km/s ; 12.2 GHz peak of 500 Jy at -50.8, 250 Jy at -51.1 km/s (Caswell 95b) ; 107 GHz peak of 12.5 Jy at -50 km/s 
{\bf G$\,\mathbf{323.740-0.263}$ :} This is a strong ($\sim$ 3000 Jy) 6.7 GHz methanol maser.  All of the observed class II methanol maser transitions in this source (including the newly detected 37.7~GHz maser) peak at velocities between -52 and -50~\kms.

% 6.7 GHz velocity range -58 - -49 km/s (MMB)
% Check the second lot of Tid 19.9 GHz observations 19.9 GHz no published data
{\bf G$\,\mathbf{337.705-0.053}$ :} This source was observed at 107 GHz by \citet{Caswell+00} who list it as possibly having weak 107 GHz methanol maser emission in the range -55 -- -50 \kms\/ because of a deviation in the thermal profile over this velocity range compared to that seen in the 156.6 GHz transition.  The 6.7 and 12.2 GHz methanol masers in this source have their peak at a velocity of -54.6~\kms\/ \citep{Caswell+11,Breen+11b}, similar to the observed 37.7 GHz emission peak at -55~\kms.  The detection of a 37.7~GHz maser in this source strengths the likelihood that the tentative 107 GHz maser detection is real (see section~\ref{sec:timeline}).

%19.9 GHz, 9.8 Jy -36.2 km/s, FWHM 0.4 km/s and 1.5 Jy, -36.8 km/s, FWHM 0.4 km/s ; 6.7 GHz peak -38.7 km/s, 12.2 GHz peak  -38.7 km/s ; 6.7 GHz velocity range -41 - -27 (MMB)
{\bf G$\,\mathbf{339.884-1.259}$ :} This is one of the strongest 6.7 GHz methanol maser sources and has been detected in a number of the weaker class II maser transitions including 107 GHz \citep{Valtts+99,Caswell+00} and 19.9 GHz \citep{Ellingsen+04}, and now in the 37.7 GHz transition.  All of these transitions, with the exception of the 19.9 GHz peak at a velocity of -38.7 \kms.  The 37.7 GHz emission has a peak flux density of approximately 320 Jy, making it the strongest known source in this transition.

% No 19.9 GHz detection $>$ 0.76 Jy, 107 GHz peak of 6.1 Jy at -105.9 km/s ; 6.7 GHz peak of about 158 Jy at -108.1 (MX), -105.1 (Cube) Caswell+11 comment that this source has a very wide velocity range and is quite variable ; 12.2 GHz peak of 42 Jy at -105.3 km/s ; 6.7 GHz velocity range -111.5 - -85 (MMB).
{\bf G$\,\mathbf{340.785-0.096}$ :} This source has been previously detected in the 6.7, 12.2 and 107 GHz transitions.  The strongest emission is most often at a velocity of around -105~\kms (the velocity of the 37.7 GHz peak), although the most recent 6.7 GHz observations of \citet{Caswell+11} find the strongest 6.7 GHz emission at -108.1~\kms.

% 19.9 GHz, 0.21 Jy, -17.6 km/s, FWHM 1.5 km/s ; 107 GHz peak of 82 Jy at -21.7 km/s ; 6.7 GHz peak at -21 km/s, total range -24 - -16 km/s ; 12.2 GHz peak at -21.8 km/s
{\bf G$\,\mathbf{345.010+1.792}$ :} This source holds the record for the largest number of class II methanol maser transitions observed towards any one source \citep[see for example][]{Cragg+01}.  Observations of this source have been made in all but one of the known class II methanol maser transitions, the exception being the 29.0 GHz transition, (for which the only published search is by \citet{Wilson+93}).  Maser emission has been detected in all of these transitions, except for the 23.1 GHz transition which was not detected in the sensitive search of \citet{Cragg+04}.  The observations undertaken here are the first towards this source in the 37.7, 38.3 and 38.5 GHz transitions, all of which are detected.  The 37.7 GHz maser in this source has a peak flux density in excess of 250 Jy (it is the second strongest source in this transition) and it is the only new detection in the 38.3 and 38.5 GHz transitions.  Early observations of this source at 6.7 GHz in 1992 \citep[e.g.][]{Caswell+95a} showed the peak at around -17~\kms, but in more recent observations \citep{Ellingsen+04,Caswell+10} that feature has declined and the strongest feature is at -21~\kms, very close to the velocity of the peak emission in the 12.2, 37.7, 38.3, 38.5 and 107 GHz transitions.  The 38.3 and 38.5 GHz spectra clearly show that there are at least two spectral features at velocities around -22 -- -21~\kms, which helps to explain some of the difficulty encountered by \citet{Cragg+01} simultaneously modelling all the methanol transitions in this region with a single component.  Interestingly the 19.9 GHz peak is at -17.6~\kms\/ (the velocity of the old 6.7 GHz peak), and there is also 12.2 GHz emission and weak 107 GHz maser at the same velocity.  However, there is no 37.7, 38.3 or 38.5 GHz emission near -17.6~\kms\/ at the sensitivity of these observations.

% No 19.9 GHz detection $>$ 0.39 Jy ; 107 GHz peak of 7.6 Jy at -3.3 km/s ; 6.7 GHz peak of 65 Jy at -3.5 km/s, velocity range -17.5 -- -2.5 km/s ; 12.2 GHz emission 34 Jy at -3.5 km/s
% Note : the peak here is at the red-shifted end of the spectrum???
{\bf G$\,\mathbf{348.703-1.043}$ :} The peak of the methanol masers at 6.7, 12.2, 107 and the newly detected 37.7 GHz maser all lie at a velocity of approximately -3.5~\kms.  At 6.7 GHz this source has a velocity range of -17.5 -- -2.5~\kms\/ \citep{Caswell+10}, which places the peak near the red-shifted extreme.

%19.9 GHz 147 Jy, -10.4 km/s, 0.3 km/s ; 26.8 Jy, -11.0 km/s, 0.3 km/s ; 1.9 Jy, -9.6 km/s, 0.4 km/s ; 6.7 GHz UC peak of 3400 Jy at -10.4 km/s, range -12 - -6, non UC peak of 1800 at -11.1, range -12 - 7 km/s ; 12.2 GHz UC peak of 976 Jy at -10.4 and non UC peak of 692 at -11.2 km/s.

{\bf NGC6334F (G$\,\mathbf{351.416+0.645}$):}  This is a very well studied southern star formation region, which shares many characteristics with the W3(OH) region.  Like W3(OH), it shows emission in a very large number of class II methanol maser transitions \citep{Cragg+01} and was one of the sources detected in the 37.7, 38.3 and 38.5 GHz search of \citet{HBM89}.  There are in fact two sites of strong ($>$ 1000 Jy peak flux density) 6.7 and 12.2 GHz methanol masers separated by around 2$\arcsec$ with overlapping velocity ranges.  One of the sites is projected against the strong cometary UCH{\sc ii} region and has its peak 6.7 GHz emission at -10.4~\kms\/ \citep{Caswell+10}, while the other is offset \citep{Ellingsen+96,Caswell97} from the radio continuum and has its 6.7 GHz peak emission at -11.1~\kms.  These two sites are far too close to distinguish in the present observations and interferometry will be required to determine if the 37.7 and 38 GHz masers are associated with both regions or just one.  However, for each of the three transitions Gaussian fitting (Tables~\ref{tab:results37}, \ref{tab:results383} \& \ref{tab:results385}) shows one spectral peak at a velocity of around -10.4~\kms\/ and a second at a velocity around -11~\kms.  The 38 GHz transitions being strongest at the former and the 37.7 GHz being strongest at the latter.  NGC6334F is the only source which shows strong emission ($>$20 Jy) in all of these three transitions.

In 1989 the 37.7 GHz emission in this source had a peak flux density of approximately 100~Jy at a velocity of -11~\kms\/ (after adjusting the velocity to the same rest frequency used here).  In the intervening 20 years the peak of the 37.7~GHz emission has declined by approximately 30\%.  The velocity of the peak emission at 38.3 and 38.5~GHz also agree to within 0.1~\kms\/ with that observed by \citet{HBM89}, but while the 38.3 GHz emission is significantly weaker (a peak of 120~Jy, compared to 330~Jy in 1988), the 38.5~GHz emission has approximately the same intensity.

%No 19.9 GHz detection $>$ 0.22 Jy.  Haschick et al. (1989) observed 35~Jy at -0.73 km/s with a velocity resolution of 0.26 km/s ; 6.7 GHz peak of 5300 Jy at 1.3 km/s, range -4.8 - 8.9 (offset source has range 5 - 7 km/s) ; 12.2 GHz peak 401 Jy at 1.4 km/s ; 107 GHz peak 22 Jy at -0.5 km/s; 85.5 GHz detection at -0.9 km/s, 1.2 Jy
% The peak being at -1.1 km/s means that it is significantly shifted towards the blue end of the spectrum c.f. all the other  transitions.
% HBM89 report a source with a peak flux density of around 35 Jy peaking at -0.73 km/s (with their assumed rest frequency).
{\bf G$\,\mathbf{9.621+0.196}$ : } This is the strongest 6.7 GHz class II methanol maser, with a peak flux density in excess of 5000 Jy, and it is also distinguished in being the best studied of the ``periodic'' methanol masers \citep{Goedhart+03,Goedhart+05,Vlemmings+09}.  The 6.7 and 12.2 GHz emission in this source peaks at a velocity of around 1.3~\kms\/, but the 37.7 GHz masers are offset from this by more than 2~\kms, and peak at -1.1~\kms.   The 37.7 GHz spectrum suggests that there might be weak (~2 Jy) emission at the peak velocity of the stronger masers, but more sensitive observations are required to confirm this.  Previous observations of the 37.7~GHz methanol masers in this source \citep{HBM89} observed slightly stronger peak flux density (30~Jy), but at the same velocity (-1.0~\kms\/ after correcting for the different rest frequency).  The peak in the 107 GHz spectrum is at -0.5~\kms\/ \citep{Caswell+00}, so it is also blue-shifted with respect to the lower frequency transition, although less so than the 37.7 GHz emission.  The velocity of the 107 GHz peak may be affected by blending with the broad thermal emission in this source, which future higher spatial resolution observations may be able to resolve.  There is a 10 Jy 107 GHz peak approximately coincident in velocity with the 6.7 and 12.2 GHz peaks.  This is also one of only 5 sources which show maser emission from the 85.5~GHz $6_{-2} \rightarrow 7_{-1}\mbox{E}$ (which is in the same transition series as the 37.7 GHz transition), and this has its peak velocity at -0.9~\kms \citep{Cragg+01}.  The 37.7 GHz peak corresponds to emission which has a peak flux density of about 20 Jy at 12.2 GHz and 80~Jy at 6.7 GHz \citep{Caswell+10,Breen+11b}.

% No 19.9 GHz detection $>$ 0.24 Jy ; 6.7 GHz peak of 77 Jy at 103 km/s (range 94 -- 113) ; 12.2 GHz peak of 17 Jy at 103.8 km/s (0.5 Jy at 103 km/s where 6.7 GHz peaks) ; 107 GHz peak 4.4 Jy at 97.2 km/s ; 37.7 GHz 2.3 Jy at 98 km/s
{\bf G$\,\mathbf{23.440-0.182}$ :} This source shows class II methanol maser emission over a velocity range from 94 -- 113~\kms\/ \citep{Caswell+95a}.  The strongest 6.7 GHz emission is at velocities around 103~\kms, and strong 12.2 GHz emission has been seen at similar velocities.  The 107 and 37.7 GHz masers in this source both have peak velocities much closer to the blue-shifted end of the spectrum, at 97.2 and 98~\kms\/ respectively \citep{Caswell+00}.  At both 6.7 and 12.2 GHz there are a number of secondary spectra features in this velocity range, which are much more stable than the spectral features at velocities greater than 100~\kms\/ \citep{Caswell+95c}.

% 6.7 GHz peak of 560 Jy at 42 km/s (range +39 -- +47 km/s)  The MMB database shows the strongest 6.7 GHz emission at about 44.5 km/s as well now.  12.2 GHz peak 109 Jy at +44.6 km/s (22 Jy at +42.3 km/s - 6.7 GHz peak).  107 GHz 24 Jy at 42 km/s ; 37.7 GHz peak of 3.2 Jy at 44.5 km/s ; c.f. HBM89 peak of ~20 Jy at 44.9 km/s (using their assumed rest frequency).
%NOTE : This is another source with the emission at the red-shifted end of the spectrum.
{\bf G$\,\mathbf{35.201-1.736}$ (W48) :} This well-studied star formation region is one of the original 37.7 GHz maser detections of \citet{HBM89}, who observed emission with a peak flux density of approximately 20~Jy towards this source.  Observations with Onsala in late 2005 detected a much lower peak flux density of 3.2 Jy, while the observations made four years later with Mopra detected no 37.7 GHz emission from this source stronger than 1.5 Jy.  Further observations were undertaken with Mopra in February 2011 to check this result and also failed to detect any 37.7 GHz emission with a 3-$\sigma$ limit of 1.1~Jy.  So in a period of 20 years the 37.7 GHz emission in this source has faded by a factor of more than 10.  The 12.2 GHz emission in this source peaks at a velocity of 44.6~\kms (the velocity of the 37.7 GHz peak), while the 6.7 and 107 GHz emission peaks at a lower velocity of 42~\kms.

\subsection{Comparison of 37.7 and 6.7 GHz masers}

Figure~\ref{fig:masers37} shows that the spectra of the 37.7 (and 38 GHz) masers are much simpler than their 6.7 GHz counterparts and typically consist of one or two spectral features.  For the majority of sources the 37.7 GHz peak coincides in velocity with the 6.7 GHz maser peak to within 0.2~\kms, and where it does not, it is always coincident with a secondary peak in the 6.7 GHz spectrum (G$\,9.621+0.196$ is a notable example).  It is clear that all of the brightest 6.7 GHz masers have an associated 37.7 GHz maser, but the relative intensity of that emission varies significantly, from a few Jy, to several hundred Jy.  Also, there are a number of relatively weak ($<$ 100 Jy peak flux density) 6.7 GHz methanol masers which have an associated 37.7 GHz maser.  The 37.7 GHz masers have velocity ranges of at most a few \kms, compared to typical ranges of $\sim$10~\kms\/ for the 6.7 GHz methanol masers they are associated with.  

In any search for rare, weak masers it is important to consider the role the sensitivity limit may have on the results.  \citet{Breen+11b} have investigated this issue in some detail for their search for 12.2 GHz methanol masers towards the MMB sample and much of the discussion in their section 5.2 and 5.3 is relevant to this search.  They identify several lines of evidence that suggest that sensitivity is not the limiting factor and here we summarise the relevant ones.  In most strong 6.7 GHz masers there are multiple secondary features with flux densities within a factor of a few of the peak intensity.  The sensitivity of our observations is such that were there 37.7 GHz emission associated with these secondary peaks we would detect it, unless the 6.7:37.7 GHz flux density ratio were always much lower for the secondary features than the main peak.  This seems unlikely given that the ratio of 6.7:12.2 GHz features within a single source shows much less scatter than in the population as a whole \citep{Breen+11b}.  These observations are consistent with the hypothesis that the total volume of gas conducive to 37.7 GHz maser emission is significantly smaller than that for the 6.7 GHz emission.  Additional evidence that sensitivity is not the primary factor in determining the number of 37.7 GHz masers we have detected in our search come from section~\ref{sec:timeline}, where we show that the 37.7 GHz masers are preferentially associated with the most luminous 6.7 GHz masers, whereas our sample contains sources with a broad range of 6.7 GHz maser luminosities

\section{Discussion}

% Ideas for discussion:
% the detections tend to have their peak at the blue-shifted end of the spectrum?  Check how well this holds by comparing the peak velocity to the entire velocity range.  This is consistent with the strong emission being projected in front of the HII region (if the masing gas is in a shell outside the ionised gas.
% I've checked this and its not true, so nothing to say.  See the plots produced in commands.r, they don't show any preference for red or blue shifted overall.

% Taking only the 25 107 GHz sample, the correlation coefficient between the 6.7 and 12.2 GHz flux density is 0.76, whereas it is only 0.60 for 6.7 and 107.  There is a closer correlation between the 12.2 and 107 GHz flux densities (0.7). The 6.7 to 37 GHz correlation is poor (0.24), although better between 12.2 and 37 GHz (0.5).  

% Looking at peak luminosities flux*dist^2 rather than flux densities, the correlation coefficient between the 6.7 and 12.2 GHz is 0.85.  It is a bit less for 6.7 and 107 (0.74 actually).  The correlation coefficient for 6.7 and 37.7 is 0.16, and 0.46 for 12.2 and 37.7 GHz, and 0.75 for the 12.2 and 107 GHz.

The search which we have undertaken here can be directly compared to the search for the $6_{-2} \rightarrow 7_{-1}\mbox{E}$ (85.5~GHz) and $7_{2} \rightarrow 6_{3}\mbox{A}^-$ (86.6 GHz) methanol maser transitions undertaken by \citet{Ellingsen+03}.  The 37.7~GHz masers are from the same transition family as the 85.5~GHz, and the 38.3~GHz masers are from the same transition family as the 86.6~GHz.  The lower frequency class~II masers in each transition family are typically stronger, so we would expect to detect all of the 85.5 and 86.6~GHz maser sources, in addition to some sources not seen in those transitions.  Broadly speaking those expectations are only partially met.   A total of 45 sites have been searched for 85.5~GHz methanol masers resulting in the detection of 5 sources \citep{Cragg+01,Sutton+01,MB02,Ellingsen+03} - specifically G$\,328.808+0.633$, G$\,345.010+1.792$, G$\,9.621+0.196$, G$\,29.95-0.02$ and DR21(OH).  These have all been observed at 37.7~GHz by us, two of these resulted in 37.7~GHz detections with a peak flux density around 20 times greater than that observed at 85.5~GHz.  However, for G$\,328.808+0.633$, G$\,29.95-0.02$ and DR21(OH), no 37.7~GHz emission was detected and we are able to place an upper limit of $\sim$1 on the 37.7:85.5 GHz peak flux density ratio.  Looking at the converse cases where we have a 37.7~GHz maser with no associated 85.5~GHz maser we can set a lower limit on the 37.7:85.5 GHz peak flux ratio in those sources and these values range from $<$1 through to approximately 300 for G$\,339.884-1.259$.  So the range of observed ratios spans at least 2.5 orders of magnitude, which is comparable to that observed for the 6.7:12.2 GHz peak flux density ratio \citep{Breen+10}.

The correspondence between the 38.3/38.5 GHz ($6_{2} \rightarrow 5_{3}\mbox{A}^-/\mbox{A}^+$) and the 86.6/86.9 GHz ($7_{2} \rightarrow 6_{3}\mbox{A}^-/\mbox{A}^+$) transitions appears generally closer, but this could be the result of small number statistics.  Two of the three 38.3/38.5 GHz sources also have associated 86.6/86.9 GHz maser emission (W3(OH) and G$\,345.101+1.792$), and in both of these cases the 38 GHz masers are approximately a factor of 2 stronger than the 86 GHz masers.  The exception is NGC6334F, which has strong ($>$100 Jy peak) 38 GHz masers, but no detected 86 GHz masers at a sensitivity of a few Jy.

\subsection{Comparison of Different class II methanol transitions}

Combining the results presented here with data from the literature we are now in a position to compare the characteristics of a dozen different class II methanol maser transitions in a sample of 25 sources.   These dozen transitions constitute all the published class II methanol maser transitions with the exception of the 29.0~GHz $8_{2} \rightarrow 9_{1}A^{-}$ \citep{Wilson+93} and the 157 GHz $J_0 \rightarrow J_{-1}E$ series \citep{Slysh+95}.  The 25 sources represent the complete sample of currently known 107~GHz methanol masers.  The data is collated in Table~\ref{tab:compare} and is approximately 95\% complete for these transitions (the least complete transition is the 108.8~GHz observations).  The comparison of the emission from the different transitions has been made at the velocity of the 37.7~GHz peak, or where there was no emission in that transition detected, at the velocity of the 107~GHz peak.  This table shows that 37.7~GHz is the fourth most common class II methanol maser transition, having been detected towards approximately half of the sample, with many of the detections having peak flux density in excess of 10 Jy.

At the bottom of the table we have attempted to summarize some of the key characteristics of the twelve transitions for which moderately large searches have been made.  Some of the numbers are a little uncertain, for example the exact number of 6.7 GHz methanol masers which have been detected keeps increasing as the results of the methanol multibeam (MMB) survey (and other searches) are published, and determining exactly how many class II sources have been searched for a particular transition is complicated by inconsistencies in naming sources and similar issues.  However, these minor uncertainties are insufficient to obscure the general trends displayed by the different transitions.  It is clear from Table~\ref{tab:compare} that the population of 6.7 and 12.2~GHz transitions differs significantly from the other class II methanol maser transitions.  The 6.7~GHz transition is the only one for which any significant unbiased searches have been undertaken and extrapolating the results from the MMB suggests that the total Galactic population exceeds 1000 sources.  A sensitive followup towards all the MMB detections south of declination -20$^\circ$ detected 12.2 GHz masers towards 43\% of 6.7 GHz methanol masers \cite{Breen+11a}.  For the remaining transitions, generally of order 50 class II maser sites have been searched, typically with a strong bias towards those with strong 6.7 GHz methanol masers.  Given this bias, the tabulated detection percentages for these remaining transitions are likely to be upper limits on the detection percentage taken over the entire population of all class II maser sites.  Aside from the 6.7 and 12.2~GHz methanol masers, it is clear that the other class II transitions are much rarer (detected towards $<$ 5\% of the 6.7~GHz population), and where they are detected, the emission is almost always weaker than the 6.7 GHz peak emission in all other transitions.  Hereafter, we use the term strong, common methanol masers to refer to the 6.7 and 12.2~GHz transitions and rare, weak to refer to the other class II methanol maser transitions.

Looking at the results summarized in Table~\ref{tab:compare} the only common pattern which can be seen is that many of the sources (9 out of 25) only show emission at 6.7, 12.2 and 107 GHz.  More generally, two thirds of the sample exhibits class II methanol maser emission in at least one other transition.  However, beyond a propensity for some transitions to be more common than others, many different combinations of rare, weak transitions are observed.  Where the rare, weak transitions are observed they are often at the same velocity as the strongest 6.7 and 12.2 GHz emission, although in some sources (e.g. G$\,9.621+0.196$), this is not the case.  Also, several of the 19.9 GHz class II masers (G$\,339.884-1.259$ and G$\,345.010+1.792$)  peak at velocities significantly offset from the strongest emission in other transitions suggesting that this transition in some cases favors different conditions from those which give rise to the presence of the rare, weak masers in most sources.

\subsection{A maser-based evolutionary timeline} \label{sec:timeline}

Assuming that all Galactic high-mass star formation regions show maser emission during their evolution, the presence and absence of different maser transitions towards specific regions can potentially be used to trace the changes in physical conditions and hence their evolutionary state.  While this is not a new idea \citep[see for example][]{Lo+75,GD77}, it requires large, sensitive, high resolution, preferably unbiased samples of a variety of maser transitions which are only now becoming available. Combining data from existing unbiased OH maser searches \citep{Caswell98} with the recently released MMB survey \citep{Caswell+10,Green+10,Caswell+11}, the HOPS 22 GHz water maser survey \citep{Walsh+11} and complementary data at other wavelengths (e.g. the Spitzer GLIMPSE Legacy survey and the submillimeter ATLASGAL survey) it is becoming feasible to make statistical studies of the properties of high-mass star formation regions associated with masers.  A qualitative maser-based evolutionary timeline involving all the strong, common maser transitions observed in high-mass star formation regions was first proposed by \citet{Ellingsen+07} and has recently been improved and quantified by \citet{Breen+10}.  High-mass star formation occurs predominantly in clusters, where there are often observed to be objects at a range of evolutionary stages within close proximity \cite[e.g.][]{Wang+11}.  However, class II methanol masers are confined to small volumes, close to a particular young stellar object within the cluster, so we expect the maser evolutionary timeline to reflect the properties of an individual object, rather than the cluster as a whole.  This is supported by the results of \citeauthor{Breen+10} which show that the masers are more sensitive probes of changes in physical condition (and hence evolutionary state) than properties such as the mid-infrared colors, which reflect the properties of the bulk emission of warm dust.  Furthermore, where multiple class II maser transitions are observed towards the same region observations to date show that they are always coincident on arcsecond scales.  This means that although interferometric observations are required to establish the coincidence definitively, we are confident that it is reasonable to use our observations with spatial resolutions of 1-2$\arcmin$ and assume that the emission from the various transitions are associated with the same individual object within the cluster environment.

\citeauthor{Breen+10} find that the luminosity of the 6.7 and 12.2 GHz class II methanol masers is largest for the most evolved regions.  This is consistent with theoretical models which predict that the luminosity will increase as the gas temperature increases, the density decreases and emission from the ultra-compact HII region increases (with the maser luminosity expected to catastrophically drop at some point, as discussed below).  The rare, weak class II methanol maser transitions are typically found towards the brightest 6.7 GHz methanol masers and so studies of these transitions towards a larger sample of sources can potentially shed light on relatively short-lived phases during the evolution of high-mass star formation regions.   It is well established that 6.7 GHz masers trace an evolutionary phase which largely precedes the formation of a UC\ionhy\/ region \citep{Walsh+03,Minier+05,Ellingsen06}.  The exact mass range of the objects with an associated 6.7 GHz methanol maser is not well established, but it is known that class II methanol masers are not associated with low-mass stars \citep{Minier+03,Xu+08} .  \citet{Ellingsen+04} investigated the radio continuum emission and OH masers associated with the 25 known 107 GHz methanol masers (see their table~3) and found 50\% show centimetre radio continuum emission stronger than a few mJy, while all but one have an associated main-line OH maser (the exception is G$\,192.600-0.048$), and 50\% have an associated OH satellite or excited-state maser.  The presence of centimetre radio continuum and/or an OH maser (particularly an excited-state OH maser) strongly suggests a relatively evolved 6.7 GHz methanol maser region and so it appears that the 107~GHz methanol maser phase likely traces a late part of the class II methanol maser phase in star formation regions. 

\begin{figure}
\includegraphics[angle=270,scale=0.7]{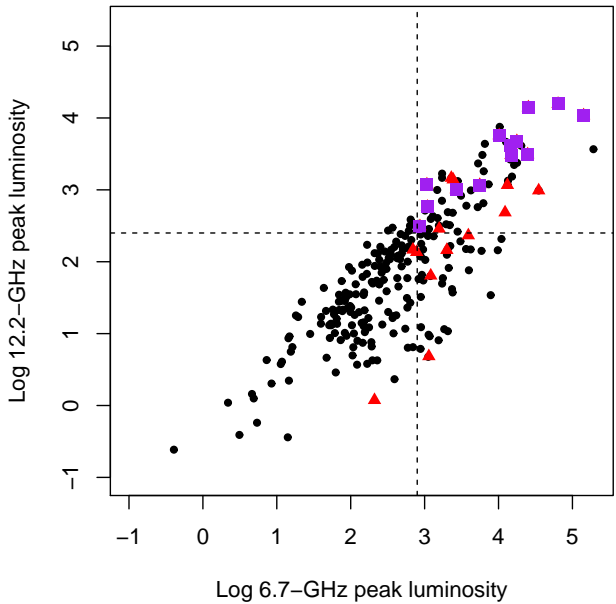}
\caption{The luminosity of the 12.2 GHz maser peak versus the luminosity of the 6.7 GHz maser peak for various maser samples.  The red triangles represent sources with an associated 107~GHz methanol maser (but no 37.7 GHz maser) and the purple squares represent sources with associated 37.7 GHz methanol masers (all of which also have an associated 107 GHz maser).  The black dots represent the 237 12.2 GHz methanol masers at declinations $<$ -20$^\circ$, from the survey of \protect\citep{Breen+11a}, The calculated luminosities assume isotropic emission and are in units of Jy kpc$^2$.  For the distances used in the luminosity calculations we have used parallax measurements and H{\sc i} self-absorption where available, and assumed the near kinematic distance for the remaining sources.  The vertical dashed line at 2.9 approximates the observed cut-off for the presence of 107 GHz masers (red triangles and purple squares), with the horizontal line at 2.4 showing the region containing all the  37.7 GHz masers (purple squares).} \label{fig:lumplot}
\end{figure}

\citet{Breen+11a} have shown that there is a correlation between the integrated luminosity of the 6.7 and 12.2 GHz methanol masers and \citet{Breen+10} showed a similar correlation using the peak luminosity.  Using the data summarised in Table~\ref{tab:compare} we investigated the relationship between the peak luminosities of various methanol maser transitions, but found that in all cases the correlation is weaker than that observed for the 6.7 and 12.2 GHz emission.  Figure~\ref{fig:lumplot} shows the luminosity of the peak 12.2 GHz methanol maser emission versus the luminosity of the peak 6.7 GHz methanol maser emission for several different samples of sources.  The black dots combine the 6.7 GHz data from the MMB and the 12.2 GHz survey of \citet{Breen+11a} and represent all 12.2 GHz maser detections south of declination -20$^\circ$.  We have in all cases used the best available estimate of the distance to calculate the luminosity, including parallax distances from \citet{Reid+09,Rygl+10}, H{\sc i} self-absorption distances from \citet{Green+McClure11} and where neither of these were available we have used the near kinematic distance calculated using the method of \cite{Reid+09}.  In total, parallax or H{\sc i} self-absorption distances were available for 84\% of the 107 GHz sample and 68\% of the comparison sample. Figure~\ref{fig:lumplot} demonstrates the well established tendency for the peak of the 12.2 GHz emission to be weaker than the 6.7 GHz emission (although with several orders of magnitude scatter in the observed ratio), and we have plotted the same luminosity range on each axis so that the relationship can clearly be seen.  The red triangles and purple squares are the 25 known 107 GHz methanol masers detected in the surveys of \citep{Valtts+95,Caswell+00} which are summarized in Table~\ref{tab:compare}, with the purple squares represent the 37.7 GHz masers detected in this work (all of which have an associated 107 GHz methanol maser ; although see the individual comments for G$\,337.705-0.053$ for clarification for this source).  These sources are largely (but not completely), a subset of the 12.2 GHz maser detections south of declination -20$^\circ$ and for these we use either the red circle, or purple square symbols (rather than a black dot).  Figure~\ref{fig:lumplot} shows that with one exception (interestingly G$\,192.600-0.048$ again), all the 107 GHz methanol masers have a 6.7 GHz peak luminosity \gta 1000 Jy kpc$^2$, which represents only a small fraction ($\sim$15\%) of the class II methanol maser population as a whole.  The largest and most sensitive search for 107 GHz methanol masers was undertaken by \citet{Caswell+00} and their sample of 84 sources deliberately included sources covering a wide range of 6.7 and 12.2 GHz peak flux densities.  This means that the observed preference for 107 GHz methanol masers to be observed towards only the most luminous 6.7 GHz masers cannot be ascribed solely to selection bias (although a far greater fraction of the luminous sources have been searched for 107 GHz than the less luminous ones).

Since the 6.7 and 107 GHz transitions are both from the same family, a relationship between 6.7 GHz luminosity and the presence of a 107 GHz maser is perhaps not surprising; however, the degree of correlation between 6.7 and 107 GHz peak flux densities is lower than between 6.7 and 12.2~GHz.  This figure also shows that the 37.7~GHz methanol masers are only present towards the most luminous 12.2 GHz methanol masers (those for which the 12.2 GHz peak luminosity is \gta 250 Jy kpc$^2$).  The close relationship between the presence of a luminous 12.2 GHz maser and 37.7 GHz masers is likely due to the physical conditions required rather than some quirk of the molecular quantum mechanics, as these two transitions are not in the same family, although both have their final state in the $K = -1$ tree of the E rotational species of methanol.

In general we would expect the integrated luminosity of the maser emission to be a more reliable indicator of the physical conditions in a region as it depends on the conditions over a larger gas volume and will be less influenced by the stochastic nature of the maser process than is the case for a single spectral peak.  We have undertaken the same analysis using the integrated 6.7 and 12.2 GHz maser luminosities and find that for the sources were data is available the same relationships are observed as seen in Figure~\ref{fig:lumplot}.  Unfortunately integrated flux densities at 6.7 and 12.2 GHz are only available in the literature for approximately two-thirds of the 107 GHz methanol maser sample, whereas peak flux densities are available for all sources.  For this reason we have chosen to show the peak luminosity relationship, rather than the equivalent plot for integrated luminosity.

% Estimated fraction of 6.7 GHz methanol masers with log(peak lum) > 2.9 = 90/237 * 0.43 = 0.16.  Minimum fraction of these sources which have a 107 GHz maser is 42% => 107 GHz phase lasts between 7 and 16% of the 6.7 GHz lifetime. Combining the lowest fraction with lower end of the 6.7 GHz lifetime and the larger with the upper we obtain a lifetime range of 1750 - 7200 yrs.
% Estimated fraction of 12.2 GHz methanol masers with log(peak lum) > 2.4 = 52/237 * 0.43 = 0.094.  Minimum fraction of these sources which have a 107 GHz maser is 53% => 37.7 GHz phase lasts between 5 and 9.4% of the 6.7 GHz lifetime. Combining the lowest fraction with lower end of the 6.7 GHz lifetime and the larger with the upper we obtain a lifetime range of 1250 - 4230 yrs.
We have looked at those 12.2 GHz methanol masers from the \citet{Breen+11a} sample for which the logarithm of their 6.7 GHz peak luminosity $>$ 2.9 and find that 90 of the 237 sources meet this criterion.  Of these 45 have been searched for 107 GHz methanol masers by \citet{Caswell+00} or \citet{Valtts+99}, with maser emission detected towards 42\% (we have counted those sources for which \citet{Caswell+00} suggested possible 107 GHz maser emission as masers for this purpose), and 107 GHz thermal emission detected towards a further 14\%. Considering only the sources which also have the logarithm of their 12.2 GHz peak luminosity $>$ 2.4 there are 52 of 237 sources meeting these combined criteria.  Of these only 15 have been searched for 37.7 GHz masers resulting in 8 detections (a detection rate of 53\%).  The class II methanol maser sources meeting the criteria outlined above which have not previously been searched for 107 and/or 37.7 GHz masers are clearly good targets for future searches in these transitions, as if similar detection rates are achieved it will result in a significant increase in the known number of sources for each of these transitions.  We can use the information on the detection rates for the 107 and 37.7 GHz masers and the fraction of the total class II maser population that lie in the region of the 6.7 - 12.2 GHz peak luminosity plot favored by each of these transitions to estimate their lifetime by bootstrapping from the lifetime estimate of 6.7 GHz methanol masers of \citet{vanderWalt05}.  Doing this we estimate the 107 GHz methanol maser phase lasts between 2 -- 7 x 10$^3$ years and we infer a lifetime of 1 -- 4 x 10$^3$ years for the 37.7 GHz transition.  Working on the assumption that the interpretation of \citet{Breen+10}, that the 6.7 and 12.2~GHz methanol maser luminosity increases as the sources evolve, is correct; the sources showing 107 GHz, and in particular 37.7 GHz methanol masers are the most evolved sources traced by class II methanol masers and arise just prior to the switch off of class II methanol masers.  

The results of \citet{Breen+10} suggest that the ``turn-off'' of class II methanol masers in star formation regions is a rapid one.  The reason for this ``turn-off'' remains an open question.  The modeling results of \citet{Cragg+05} suggest that the presence of the rare weak masers in more evolved sources is indicative of the physical conditions evolving toward those which favor a large number of methanol transitions.  The sudden disappearance of the class II methanol masers, which reside very close to the protostar, suggests a catastrophic change in the environment of the masing gas.  Possibilities include the conditions reaching a critical point where the methanol is rapidly depleted by gas-phase reactions, destruction of methanol by increased UV photon flux, or disruption of the velocity coherence in the masing gas by the passage of a shock front.  Identifying these rare ``post-peak-luminosity'' class II methanol maser sources is likely to give us new insights into some of the critical factors governing the presence or otherwise of the masers.  Figure~\ref{fig:lumplot} identifies one possible candidate for a post-peak-luminosity maser, in the form of the outlying 107 GHz methanol maser source G$\,192.600-0.048$.  This source is significantly less luminous at 12.2 GHz than the bulk of 107 GHz maser sources and also the least luminous at 6.7 GHz.  Were it not that this source is relatively nearby (it is at a distance of 1.59~kpc, measured to an accuracy of better than 5\% using parallax by \citet{Rygl+10}), we would only be able to detect the 6.7 GHz methanol masers.  Other candidates for post-peak-luminosity class II maser sources would be objects which have strong HII regions and low density (or high temperature), accompanied by a maser with a peak luminosity \lta 1000 Jy pc$^2$.

\section{Conclusions}

We have undertaken a search for 37.7, 38.3 and 38.5~GHz class II methanol masers towards a large sample of 6.7 GHz methanol masers covering both the northern and southern hemispheres.  This led to the detection of thirteen 37.7 GHz methanol masers, eight of which are new detections and the detection of three 38.3/38.5 GHz methanol masers, one of which is a new detection.  We find that 37.7 GHz methanol masers are only detected towards the class II methanol maser sources which have the highest 6.7 and 12.2 GHz peak luminosities.  In the developing maser-based evolutionary timeline, these sources are thought to signpost the latter stages of the class II methanol maser phase in star formation regions.  We estimate that 37.7 GHz methanol masers have a lifetime of only a few thousand years and arise just prior to the changes which terminate methanol maser emission in the region, i.e. they are the horsemen of the apocalypse for the class II methanol maser phase.

\section*{Acknowledgements}

%\acknowledgements 
SPE would like to thank the Alexander-von-Humboldt-Stiftung for an
Experienced Researcher Fellowship which has helped support this research.
AMS  was  partially supported by the Russian Foundation for Basic Research (grant nos. 10-02-00589,
11-02-01332 and 11-02-97124).  We are grateful to \framebox{L.E.B. Johansson} and A.I. Vasyunin for help with the Onsala
observations.  The Mopra telescope is part of the Australia Telescope which is funded
by the Commonwealth of Australia for operation as a National Facility
managed by CSIRO. The OSO is the Swedish National Facility for Radio Astronomy and is operated by Chalmers University of Technology, G\"oteborg, Sweden, with Þnancial support from the Swedish Research Council and the Swedish Board for Technical Development.  This research has made use of NASA's Astrophysics Data 
System Abstract Service.

\clearpage

\begin{sidewaystable}
  \caption{The flux density at the velocity of the peak in the 37.7 GHz 
    transition, (or the 107.0 GHz transition for sources with no 37.7 GHz 
    emission) of the 6.7, 12.2, 19.9, 23.1, 37.7, 38.3, 38.5 85.5, 86.6, 107.0, 
    108.8 \& 156.6 GHz methanol transitions towards all star formation 
    regions with known 107.0 GHz methanol masers.  The 37.7, 38.3 and 38.5 GHz
    data is taken from this work. 
    The rest of the information has been taken from the 
    literature references given in the last column.  Where upper limits are 
    quoted they are 3 times the RMS noise level in the spectra.  
    Transitions where there is maser emission, but none at a velocity corresponding to 
    the 37.7 GHz peak are indicated with a, $^\dagger$.
    Transitions where thermal emission is detected are indicated with a $^*$
    and the upper limited listed is 50\% of the flux density of the thermal 
    emission at the listed velocity.  The total number of detections in each transition
    are tabulated at the bottom of the table, the numbers in brackets give the total number
    of detections from that transition at all velocities, not just the single velocity for each source
    listed in the table.  The number of sources detected (given in the second last line), refers
    to all maser detections, not just those from the sample of 25 sources presented here and the
    percentage detection is with respect to the number of class II methanol maser positions
    searched.  The median flux density refers only to the sample of 25 107 GHz methanol masers.
    References : 
    1=Batrla \etal\/ (1987);
    2=Breen \etal\/ (2010);
    3=Breen \etal\/ (2011b);
    4=Caswell \etal\/ (1995a);
    5=Caswell \etal\/ (1995b);
    6=Caswell \etal\/ (2000);
    7=Caswell \etal\/ (2010);
    8=Caswell \etal\/ (2011);
    9=Cragg \etal\/ (2001);
    10=Cragg \etal\/ (2004);
    11=Ellingsen \etal\/ (2003);
    12=Ellingsen \etal\/ (2004);
    13=Green \etal\/ (2010);
    14=Koo \etal\/ (1988);
    15=Mehringer, Zhou \& Dickel (1997);
    16=Menten (1991);
    17=Minier \& Booth (2002);
    18=Slysh, Kalenskii \& Val'tts (1995);
    19=Sutton \etal\/ (2001);
    20=Szymczak, Hrynek \& Kus (2000);
    21=Val'tts \etal\/ (1995);
    22=Val'tts \etal\/ (1999);
    23=Wilson \etal\/ (1984);
    24=Wilson \etal\/ (1985)
  } \label{tab:compare}
  	  \tiny
  \begin{tabular}{lrrrrrrrrrrrrrl}
                 &                & \multicolumn{10}{c}{{\bf Flux Density}}
                                              &                  \\
    {\bf Source} & {\bf Velocity} & {\bf 6.7 GHz}      & {\bf 12.2 GHz}     & 
      {\bf 19.9 GHz}    & {\bf 23.1 GHz}     & {\bf 37.7 GHz}     & {\bf 38.3 GHz}     & 
      {\bf 38.5 GHz}    & {\bf 85.5 GHz}     & {\bf 86.6 GHz}     & {\bf 107.0 GHz}   & 
      {\bf 108.8 GHz}  & {\bf 156.6 GHz}  & {\bf references} \\
                          & {\bf (\kms)}     & {\bf (Jy)}              & {\bf (Jy)}                & 
      {\bf (Jy)}               & {\bf (Jy)}               & {\bf (Jy)}                & {\bf (Jy)}                & 
      {\bf (Jy)}               & {\bf (Jy)}               & {\bf (Jy)}                & {\bf (Jy)}                & 
      {\bf (Jy)}               & {\bf (Jy)}               &                             \\ \hline
  W3(OH)          &  -43.0 & 3000 & 600    & 44.3  & 9.5            & 2.2  & 11.0      & 16.0       &
    $<$0.7  & 6.7   & 72   & $<$0.6     & $<$9     & 16,18,19,23,24 \\
  G$\,188.946+0.886$   &   10.9 & 495  & 235    & $<$0.20  &         & 23.4      & $<$1.8  & $<$3.3  &
    $<$2.1  & $<$0.8  & 15.5 & $<$4.8     & $<$2     & 4,5,6,9,11,12,22 \\
  G$\,192.600-0.048$   & 4.2    & 72   & $<$0.4$^\dagger$ & $<$0.19  &          & $<$2.4  & $<$2.1  &               & 
    $<$1.8  & $<$2.2  & 5.8  & $<$4.5     & $<$3     & 2,4,6,11,12,22 \\
  G$\,310.144+0.760$   & -56    & 130  & 114    & $<$0.16  & $<$0.8  & $<$1.2  & $<$1.2  & $<$1.2  &
    $<$1.4  & $<$2.1  & 23   &           & $<$3     & 6,10,11,12 \\ 
  G$\,318.948-0.196$  & -34.2  & 780  & 180    & $<$0.17  & $<$0.8  & 9.3        & $<$ 1.5  & $<$1.2  &
    $<$2.0  & $<$2.2  & 5.7  & $<3^{*}$   & 2.4      & 4,5,6,10,11,12,22 \\
  G$\,323.740-0.263$  & -51.2  & 132  & 135    & 0.15     & $<$0.9      & 14.6      & $<$1.2  & $<$ 1.2  &
    $<$1.8  & $<$1.9  & 9.5 & $<$5.4    & $<$4$^*$ & 6,10,11,12,22 \\ % 156 GHz thermal 6.2$^*$
  G$\,327.120+0.511$   & -89.8  & 25   & 5      & $<$0.17  & $<$0.9     & $<$1.2  & $<$1.2  & $<$1.2  &
    $<$1.5  & $<$2.2  & 9.2  &           & $<$2.8   & 4,5,6,10,11,12 \\
  G$\,328.808+0.633$   & -43.5  & 315  & 5      & 0.8      & $<$1.0         & $<$1.5  & $<$1.5  & $<$1.5  &
    1.6         & $<$2.4  & 5.5  & $<6^{*}$  & $<8^{*}$ & 6,9,10,11,12,22 \\ % 156 GHz thermal 25$^*$ 
  G$\,336.018-0.827$  & -40.2  & 30   & 7.7     & $<$0.21  & $<$0.8      & $<$1.8 & $<$1.8  & $<$1.8  &
    $<$1.3  & $<2.7$  & 6    & $<$5.7    & $<$5$^*$ & 3,6,8,9,10,11,12,22 \\ % 156 GHz thermal 3$^*$, 
  G$\,339.884-1.259$  & -38.7  & 1520 & 846    & $<$0.2$^\dagger$ & $<$0.6       & 323        & $<$2.4   & $<$2.4  &
    $<$1.1  & $<1.7$  & 64   & $<$5.4    & 4        & 3,6,8,9,10,11,12,22 \\
  % Get 12.2 GHz limit from Shari.
  G$\,340.054-0.244$  & -62.8  & 1.5  & $<$0.3$^\dagger$ & $<$0.20  & $<$0.8  & $<$1.8  & $<$ 2.1  & $<$1.8  &
                   & $<$2.3  & 2.9 &           & $<$2     & 3,6,8,10,11,12 \\
  G$\,340.785-0.096$  & -105.5 & 120  & 42     & $<$0.76  & $<$0.7  & 20.5       & $<$2.1   & $<$1.8  &
    $<$1.8  & $<$1.9  & 6.1     &        & $<$2     & 3,6,8,10,11,12 \\
  % Get 12.2 GHz limit from Shari.
  G$\,345.003-0.223$  & -26.9  & 102   & $<$0.5$^\dagger$ & 0.47     & $<$0.6     & $<$2.1   & $<$2.1  & $<$2.1  &
    $<$2.0  & $<$2.9  & 3.5     & $<1.5^{*}$ & $<$6$^*$ & 3,6,7,9,10,7,22 \\ % 156 GHz thermal 12.4$^*$ 
  % Shari - what epoch is the 6.7 GHz MX shown in Figure 5 in 12 GHz paper, it seems different from the MMB paper Caswell et al. (2010)
  G$\,345.010+1.792$   & -22.1  & 200  & 200    & $<$0.2$^\dagger$ & $<$0.5     & 207         & 9.4         & 5.0         &
    10.0       & 2.8         & 4.1     & 82          & 18              & 3,6,7,9,10,11,12,22 \\
  G$\,345.504+0.348$   & -17.7  & 300    & 8.4    & $<$0.65  & $<$0.8  & $<$2.1    & $<$2.1   & $<$2.1  &
    $<$4.7  & $<$2.4  & 2.3  & $<2^{*}$  & $<$4$^*$ & 3,6,7,10,11,12 \\ % 156 GHz thermal 7.7$^*$ 
  G$\,348.703-1.043$  & -3.3   & 65   & 34          & $<$0.39  & $<$0.9  & 4.4           & $<$2.1   & $<$2.1  &
    $<$1.5  & $<$2.7  & 7.6  &           & $<$1.5   & 3,6,7,10,11,12 \\
  NGC6334F                  & -11.1  & 1840   & 692    & 30           & 4             & 70.4        & 32.9        & 44        &
    $<$2.5$^*$& $<$1.0$^*$ & 10     &          & $<$17$^*$ & 3,6,7,9,10,7,12,22 \\ %156 GHz thermal 43.8$^*$ 
  G$\,353.410-0.360$  & -20.5  & 116  & 21     & 0.33     & $<$0.8          & $<$1.8   & $<$1.8   & $<$1.8  &
    $<$2.1  & $<$2.0  & 4.5  &           & $<$2     & 3,6,7,10,11,12 \\
  % Get 12.2 GHz flux at -1.1 from Shari
  G$\,9.621+0.196$     & -1.1   & 80   & 20         & $<$0.22  & $<$0.6       & 23.6        & $<$1.8  & $<$1.8  &
    1.2         & $<$2.3  & 22   & $<3^{*}$  & $<$3$^*$ & 3,6,9,10,11,12,13 \\ % 156 GHz thermal 8.7$^*$
  G$\,12.909-0.260$   & 39.5   & 269  & 11.5   & $<$0.27  & $<$0.8    & $<$1.2   & $<$1.2  & $<$1.2  &
    $<$3.8  & $<$8.6  & 5.5  & $<3^{*}$   & $<$3$^*$ & 5,6,9,10,11,12,13,22 \\ % 156 GHz thermal 6.2$^*$
  G$\,23.010-0.411$   & 75.9   & 405  & 28     & $<$0.26  & $<$0.9     & $<$0.9  & $<$0.9  & $<$0.9  &
    $<$3.8  &                & 5.2  & $<$5.1     & $<$2     & 4,5,6,9,10,12,22 \\ % NOTE: called 23.01+0.41 in Val'tts et al. 1999
  G$\,23.440-0.182$   & 98.0   & 23   & 9      & $<$0.24  & $<$0.9        & 2.3         & $<$1.2  & $<$1.2  &
    $<$4.0$^*$ & $<$4.1  & 4    &            & $<$2     & 4,5,6,10,11,12 \\
  G$\,35.201-1.736$   & 44.5   & 260  & 109    & $<$0.51  & $<$0.9       & 3.2         & $<$1.5  & $<$1.5  &
    $<$1.1   & $<$2.3  & 22   & $<$4.8     & 4.6      & 4,5,6,9,10,11,12 \\ 
  Cep A           & -2.2   & 1420 & $<$7$^\dagger$   &                    &                     & $<$2.1  & $<$1.5  &                &
   $<$5.0   & $<$5.0  & 16   & $<$5.0     & $<$3     & 14,15,16,17,18 \\ 
  NGC7538         & -59.0  & 15   & $<$5$^\dagger$   & 0.21      & 0.21            & $<$2.1  & $<$1.8  &                &
    $<$5.0   &$<$5.0  & 10.1   & $<4.0^*$ &            & 1,20,17,21,23,24 \\ \hline
  {\bf \# Det. in Sample of 25}      & & 25                  & 20(24)         & 5(7) & 3    & 12       & 3   & 3    & 3       & 2    & 25 & 1 & 4 \\
  {\bf \# Class II searched} & & $>$800         & $>$600       & $\sim$29    & $\sim$75  & $\sim$79      & $\sim$76 & $\sim$60  & $\sim$45    & $\sim$32  & $\sim$175 & $\sim$45 & $\sim$67 & \\
  {\bf Detected \#/\%}          & & $>$800/100 & $>$260/43 & 7/24 & 3/4 & 13/16 & 3/4& 3/5 & 5/11 & 3/9 & 25/14 & 2/4 & 4/6 & \\
  % Should I do the 107 GHz sample median and the population median?
  {\bf Median peak flux (Jy)} & & 300                  & 75.5                 & 0.47         & 4       & 17.6           & 11 & 16 & 1.6 & 4.75 & 6.1 & 82 & 4.3 & \\
  \end{tabular}
\end{sidewaystable}

\end{document}